\documentclass[11pt]{article}
\usepackage{jheppub}
\usepackage{graphicx}
\usepackage{amsmath,amssymb}
\usepackage{array}
\usepackage{comment}
\usepackage{hyperref}
\usepackage{color}
\usepackage{pb-diagram}
\usepackage{slashed}
\usepackage{cancel}
\usepackage{float}
\usepackage[normalem]{ulem}
\newcommand{\be}{\begin{equation}}
\newcommand{\ee}{\end{equation}}
\newcommand{\bea}{\begin{eqnarray}}
\newcommand{\eea}{\end{eqnarray}}
\newcommand{\MPl}{M_\mathrm{Pl}}

\newcommand{\doublet}[2]{ \left( \begin{array}{c}#1 \\ [1ex] #2 \end{array}\right) }

\newcommand{\At}{\widetilde{A}}
\newcommand{\xit}{\widetilde{\xi}}
\newcommand{\Vt}{\widetilde{V}}
\def\nn{\nonumber}

\usepackage{color}

\begin{document} 
\title{\hfill ~ \\[-40mm]
\begin{footnotesize}
\hspace{130mm}
\normalfont{DIAS-STP-24-29}\\
\end{footnotesize}
\vspace{30mm}
\boldmath  Baryogenesis from Primordial CP Violation}
\author[\,a,b]{Venus Keus}
\author[\,c]{and Edward W. Kolb}

\affiliation[a]{School of Theoretical Physics, Dublin Institute for Advanced Studies, 10 Burlington Road, Dublin, D04 C932, Ireland}
\affiliation[b]{Department of Physics and Helsinki Institute of Physics, Gustaf Hallstromin katu 2, FIN-00014, University of Helsinki, Finland}
\affiliation[c]{Kavli Institute for Cosmological Physics and Enrico Fermi Institute, 
The University of Chicago,  5640 South Ellis Avenue, Chicago, IL 60637, U.S.A.}

\emailAdd{venus@stp.dias.ie}
\emailAdd{Rocky.Kolb@uchicago.edu}

\abstract{
We present a novel Baryogenesis mechanism in which an asymmetry of scalars in a three-Higgs doublet model produced exiting a CP-violating inflationary set-up is translated into an asymmetry of baryons through electroweak instantons. 
\\ \\ \\ \today
}

\maketitle

\section{Introduction}
\label{sec:intro}

The Standard Model (SM) of particle physics has been exhaustively tested and is in remarkable agreement with experimental data, with its last missing piece – the Higgs boson – discovered in 2012 by the ATLAS and CMS experiments at the CERN Large Hadron Collider (LHC) \cite{ATLAS:2012yve,CMS:2012qbp}.  Although the observed properties of the Higgs are in agreement with those of the SM, the Higgs boson may be just the first member of an extended scalar sector.  Even though so far no signs of new physics have been detected, it is well understood that the SM of particle physics is incomplete, and enlargement of the minimal Higgs sector is a promising avenue to search for physics Beyond the Standard Model (BSM). 

Although the minimal SM accounts for all LHC data, it falls short of explaining several aspects of nature, such as providing a viable candidate for Dark Matter (DM), having a mechanism to generate the baryon asymmetry of the Universe (BAU), and, unless the Higgs has a very large coupling to the Ricci scalar, lacking a candidate to drive inflation.  

These are some of the theoretical and empirical reasons why it is widely accepted that one needs to consider BSM frameworks in pursuit of a more complete theory of nature. A common characteristic of many such BSM scenarios is an extended scalar sector.  Non-minimal Higgs frameworks with conserved discrete symmetries could naturally accommodate stabilised DM candidates~\cite{Ivanov:2012hc,Ma:2007gq,Batell:2010bp,Belanger:2011ww,Adulpravitchai:2011ei,Belanger:2022esk,Jurciukonis:2022oru,Aranda:2019vda,Keus:2013hya,Yaguna:2019cvp,Hirsch:2010ru}. The scalar potential -~the least constrained sector in the SM~- if extended, could also provide a strong first-order electroweak phase transition and sufficient amount of CP violation to accommodate a successful electroweak baryogenesis mechanism~\cite{Morrissey:2012db,Kuzmin:1985mm,Turok:1990in,Cohen:1990it}. Such a framework would predict new scalars which directly couple to the SM Higgs boson and are around the electroweak scale, and therefore accessible by the LHC. This is only possible in non-minimal Higgs frameworks larger than 2-Higgs doublet model (2HDM)~\cite{Keus:2019szx}. Moreover, extended scalar sectors could offer inflaton candidates~\cite{Gong:2012ri,Choubey:2017hsq} whose positive contribution to the running of the Higgs self-coupling stabilizes the electroweak vacuum~\cite{Degrassi:2012ry,Gonderinger:2012rd,Elias-Miro:2012eoi,DiChiara:2014wha}.

In this paper we explore an extension of the minimal Higgs sector, introducing a second and third (\textit{non c'\`{e} due senza tre}) Higgs doublet, which admits the possibility of providing an inflaton to drive inflation and a mechanism to generate the observed matter-antimatter asymmetry in the Universe. 

Scalars with non-minimal couplings to gravity which are light (here, light means mass smaller than the expansion rate during inflation) are well-motivated inflaton candidates and can drive the inflation process in the early Universe, such as in the Higgs-inflation model where the SM-Higgs plays the role of the inflaton, albeit with a very large nonminimal coupling to the Ricci scalar~\cite{Bezrukov:2007ep} (for a recent review including many references to the expansive literature, see Ref.~\cite{Rubio:2018ogq}).  Another approach is to augment the minimal Higgs sector by the introduction of a singlet scalar as in $s$-inflation models \cite{Enqvist:2014zqa}.

The minimal nature of Higgs-inflation makes it a compelling inflation model. However, it suffers from the so-called \textit{unitarity} or \textit{naturalness} problem.  This is due to the fact that in order to account for the Cosmic Microwave Background (CMB) observations, the model requires an  \textit{unnaturally} large coupling of Higgs to gravity \cite{Bezrukov:2007ep}. Here, we will not comment on this issue, but just note that in the three Higgs doublet model (3HDM) there need not be unnaturally large couplings to gravity, but the quartic couplings in the scalar potential have to be very small.

With regard to the role of scalars in generation of the BAU, extensive studies have been carried out in one additional scalar singlet or in one additional scalar doublet extensions of the SM (see e.g.,~\cite{Englert:2011yb,Branco:2011iw} and references therein). These models, however, by construction, can only partly provide a solution to the SM drawbacks. One has to go beyond simple scalar extensions of the SM to incorporate both CP violation and DM into the model \cite{Keus:2019szx,Keus:2016orl}. Specifically, if CP violation is embedded in the extended dark sector, a novel phenomenon introduced for the first time in Ref.~\cite{Cordero-Cid:2016krd} and studied further in Refs.~\cite{Cordero:2017owj,Cordero-Cid:2018man,Cordero-Cid:2020yba,Keus:2020ooy}, then there will be no contributions to the Electric Dipole Moments (EDMs) and no limit on the amount of CP violation in the model since the dark sector is protected from directly coupling to the SM particles. As a result, one can construct a CP-violating DM model with unbounded dark CP violation. In fact, the CP violating dark particles need not have a Higgs-DM coupling and can interact with the SM merely through the gauge bosons, ridding the model of all current (in)direct detection and LHC bounds, while yielding relic abundance in agreement with observation through the freeze-out or freeze-in mechanisms \cite{Keus:2019szx}.

Going beyond the minimal scalar extensions of the SM, one can also allow for exotic inflationary dynamics.  Here, we introduce a novel mechanism for the BAU, namely baryogenesis from (primordial) scalar asymmetries, which originated from CP-violating inflation as proposed in \cite{Keus:2021dti}. In this mechanism, an excess of matter over antimatter is produced when the inflaton (with CP-violating couplings to gravity and SM Higgs) produces an asymmetry in the light Higgs doublet field, which then is converted to a baryon excess through inflaton interactions.

We present this novel mechanism in the context of a $Z_2$ symmetric 3HDM with a CP-violating extended dark sector \cite{Cordero-Cid:2016krd,Keus:2016orl,  Cordero:2017owj,Cordero-Cid:2018man,Cordero-Cid:2020yba,Keus:2020ooy,Keus:2021dti}. It is important to note that such a scenario will require at least three Higgs doublets, in order to allow for a complex coupling of the inflation to the Ricci scalar (in addition to complex couplings in the scalar potential). The $Z_2$ symmetry, which distinguishes the two doublets responsible for driving inflation from the SM-like Higgs doublet, forbids  Flavour Changing Neutral Currents (FCNCs). Moreover, the primordial (or `the inflationary') CP violation is introduced only through the complex couplings of the two inflaton doublets to each other and to the Ricci scalar.  Therefore, this model is in perfect agreement with all low energy experimental constraints, including EDM limits. This is not the case for a 2HDM model in which the $Z_2$ symmetry required for suppression of FCNCs will also forbid a complex coupling to the Ricci scalar. 

The paper is organized as follows: In Sec.~\ref{sec:potential} we construct the inflationary potential and discuss the conformal transformation between the Jordan and Einstein frames. In Sec.~\ref{sec:slow-roll} we derive the slow-roll parameters and show that our model is in perfect agreement with the latest CMB measurements and can accommodate a conformal value for the non-minimal inflaton-gravity coupling. In Sec.~\ref{sec:baryogenesis} we calculate scalar asymmetries resulting from the CP-violating inflation and show how they lead to an asymmetry of matter over antimatter. Finally, in Sec.~\ref{sec:conclusions} we conclude and discuss the future prospects of our framework.

\section{The scalar potential}
\label{sec:potential}

\subsection{General definitions}

The scalar potential of a 3HDM, invariant under a subgroup $G$ of phase transformations, can be decomposed into two parts: $V_0$, which respects all phase rotations, and $V_G$, which remains invariant under the subgroup $G$ \cite{Ivanov:2011ae,Keus:2013hya}. Consequently, the scalar potential of a 3HDM with $Z_2$-symmetry takes the following form:\footnote{Additional $Z_2$-symmetric terms, such as $(\phi^\dagger\phi_1)(\phi_2^\dagger\phi)$, $(\phi_1^\dagger\phi_2)(\phi^\dagger\phi)$, $(\phi_1^\dagger\phi_2)(\phi_1^\dagger\phi_1)$ and $(\phi_1^\dagger\phi_2)(\phi_2^\dagger\phi_2)$ are not included since they do not affect the physical implications of the model \cite{Cordero-Cid:2018man}.}
\bea
\label{eq:V0-3HDM}
V&=&V_0+V_{Z_2}, \nonumber \\
V_0 &=& - \mu^2_{1} (\phi_1^\dagger \phi_1) -\mu^2_2 (\phi_2^\dagger \phi_2) - \mu^2_3(\phi^\dagger \phi) \nonumber\\
&&+ \lambda_{11} (\phi_1^\dagger \phi_1)^2+ \lambda_{22} (\phi_2^\dagger \phi_2)^2  + \lambda_{33} (\phi^\dagger \phi)^2 \nonumber\\
&& + \lambda_{12}  (\phi_1^\dagger \phi_1)(\phi_2^\dagger \phi_2)
 + \lambda_{23}  (\phi_2^\dagger \phi_2)(\phi^\dagger \phi) + \lambda_{31} (\phi^\dagger \phi)(\phi_1^\dagger \phi_1) \nonumber\\
&& + \lambda'_{12} (\phi_1^\dagger \phi_2)(\phi_2^\dagger \phi_1)
 + \lambda'_{23} (\phi_2^\dagger \phi)(\phi^\dagger \phi_2) + \lambda'_{31} (\phi^\dagger \phi_1)(\phi_1^\dagger \phi),  \nonumber\\
 V_{Z_2} &=& -\mu^2_{12}(\phi_1^\dagger\phi_2)+  \lambda_{1}(\phi_1^\dagger\phi_2)^2 + \lambda_2(\phi_2^\dagger\phi)^2 + \lambda_3(\phi^\dagger\phi_1)^2  + h.c. \,,
\eea
where the three Higgs doublets, $\phi_{1},\phi_2,\phi$, transform under the $Z_2$ group, respectively, as
\be
\label{eq:generator}
g_{Z_2}=  \mathrm{\rm diag}\left(-1, -1, +1 \right).
\ee

By construction, the parameters in the $V_0$ part of the potential are real. However, parameters within the $V_{Z_2}$ sector are allowed to take complex values. The following notation will be employed consistently throughout this paper:
\be
\label{eq:complex-params}
\lambda_{j} = |\lambda_{j}| \, e^{i\, \theta_{j}} \quad (j = 1,2,3), \quad
\quad
\mbox{and}
\quad
\mu^2_{12} = |\mu^2_{12}| \, e^{i\, \theta_{12}}\,.
\ee
The composition of the doublets is as follows:
\be
\phi_1= \doublet{ h^+_1}{\dfrac{h_1+i\eta_1}{\sqrt{2}}},\quad
\phi_2= \doublet{ h^+_2}{\dfrac{h_2+i\eta_2}{\sqrt{2}}}, \quad
\phi= \doublet{G^+}{\dfrac{v+H+iG^0}{\sqrt{2}}}.
\label{explicit-fields}
\ee
The doublets $\phi_1$ and $\phi_2$ correspond to the two $Z_2$-odd doublets, which do not acquire vacuum expectation values (VEVs), i.e., $\langle \phi_1 \rangle = \langle \phi_2 \rangle = 0$, hence are referred to as \textit{inert} doublets. Meanwhile, $\phi$ is the $Z_2$-even doublet that at low energy acquires a VEV given by $\langle \phi \rangle = v / \sqrt{2} \neq 0$, and is referred to as the \textit{active} doublet. The active doublet serves as the SM Higgs doublet, where the scalar field $H$ is identified as the SM Higgs boson, and the components $G^\pm$ and $G^0$ correspond to the would-be Goldstone bosons. As indicated by the $Z_2$ transformation defined in Eq.~\eqref{eq:generator}, the vacuum configuration $(0, 0, v / \sqrt{2})$ preserves the symmetry of the potential.

In this work, we explore a scenario where the components of the inert doublets drive inflation and reheat the Universe after inflation through production of $H$, $G^0$, and $G^\pm$ through the production of $\phi^\pm$ and $\phi^0$, the charged and neutral fields within the active doublet, which, in turn, populate all SM particles in the primordial soup.  During the inflationary epoch, the VEV of the active doublet is assumed to be zero until electroweak symmetry breaking (EWSB).

CP violation is incorporated only within the inert sector\footnote{By which we mean that the neutral CP-even and CP-odds fields from the inert doublets will mix, resulting in CP-mixed neutral states. Although the quartic couplings of the inert doublets with the Higgs are complex, there is no mass mixing of the inert CP-mixed states and the SM-like Higgs.} through the complex (quadratic and quartic) couplings of the inert doublets to each other, through complex quartic couplings to the active doublet and to the Ricci scalar, which will not affect the (tree-level) interactions of the active doublet, i.e., the SM-like Higgs with the rest of the SM  due to the conserved $Z_2$ symmetry\footnote{As doublets under the SU(2) gauge group of the SM, naturally, the inert doublets couple to the $Z$ and $W$ bosons. Due to the CP-mixed nature of these inert states, their couplings to the gauge bosons will also be CP violating.}. As a result, the model is not constrained by EDM bounds \cite{Cordero-Cid:2016krd}. The lightest CP-mixed neutral state from the inert doublets is stable under the unbroken $Z_2$ symmetry, making it a viable DM candidate. However, the focus of this paper is on the inflationary behaviour of the model, and we leave a detailed discussion of its DM implications for future work.

\subsection{Inflaton potential for the 3HDM }
\label{sec:Ip3HDM}

The action of the model we consider may be expressed in the Jordan frame as:
\bea
\label{eq:action-Jordan-1}
S_J &=& \int d^4x \sqrt{-g}
\biggl[  \frac{1}{2} \MPl^2 R - D_\mu \phi_1^\dagger  D^\mu \phi_1
- D_\mu \phi_2^\dagger  D^\mu \phi_2
- D_\mu \phi^\dagger  D^\mu \phi - V(\phi_1, \phi_2,\phi) ~~~~~~~~~~~~
\nonumber \\
&& \hspace{38mm}
+ \bigl(\xi_1 |\phi_1|^2  +\xi_2 |\phi_2|^2 +\xi_3 |\phi|^2
+ \xi_4 (\phi^\dagger_1 \phi_2)  +\xi_4^* (\phi^\dagger_2 \phi_1 ) \bigr)R
\biggr], 
\eea
where $R$ is the Ricci scalar, $\MPl$ is the reduced Planck mass and the parameters $\xi_i$ are dimensionless couplings of the scalar doublets to gravity.  Note that, in principle, $\xi_4$ could be a complex parameter for which we use the notation
\be
\label{eq:xi4-phase}
\xi_4 = |\xi_4|\, e^{i\theta_4}\,.
\ee

In Eq.~\eqref{eq:action-Jordan-1} the covariant derivative, $D_\mu$, contains couplings of the scalars with the gauge bosons. However, for the dynamics during the inflation, the covariant derivative is reduced to the normal derivative $D_\mu\rightarrow \partial_\mu$. 

We assume that the energy density associated with $\phi$ is sub-dominant during inflation.  Since the charged scalers in the inert doublets do not affect the inflationary dynamics we write 
\be
\phi_1= \frac{1}{\sqrt{2}} \left(\begin{array}{c}
0 \\[2mm]
h_1 +i \eta_1
\end{array} \right)
,\quad
\phi_2= \frac{1}{\sqrt{2}} \left(\begin{array}{c}
0 \\[2mm]
h_2 + i \eta_2
\end{array} \right)
\ .
\label{eq:explicit-fields}
\ee

The part of the potential relevant for inflation is
\bea
\label{eq:approxscalarpot}
V &=& - \mu^2_{1} (\phi_1^\dagger \phi_1) -\mu^2_2 (\phi_2^\dagger \phi_2)
+ \lambda_{11} (\phi_1^\dagger \phi_1)^2+ \lambda_{22} (\phi_2^\dagger \phi_2)^2 \\
&&~ + \lambda_{12}  (\phi_1^\dagger \phi_1)(\phi_2^\dagger \phi_2)
 + \lambda'_{12} (\phi_1^\dagger \phi_2)(\phi_2^\dagger \phi_1)
 -\mu^2_{12}(\phi_1^\dagger\phi_2)+  \lambda_{1}(\phi_1^\dagger\phi_2)^2 + h.c. \nonumber
\eea
With only (the neutral components of) $\phi_1$ and $\phi_2$ relevant during inflation, one can expand the potential $V$ in Eq.~\eqref{eq:approxscalarpot} in terms of $h_1,\, h_2,\,\eta_1, \,\eta_2$.  Note that the only terms in which $\eta_1$ and $\eta_2$ fields appear are the ones with $\mu_{12}^2$ and $\lambda_1$. Expanding these terms explicitly yields $ V \supset -\mu_{12}^2 \, h_1 h_2 \, e^{i(\eta_2-\eta_1)} + \lambda_1 \, h_1^2 h_2^2 \, e^{2i(\eta_2-\eta_1)} + h.c.$  So the fields appearing in the potential are not $\eta_1$ and $\eta_2$ individually, but rather the combination $\eta_2-\eta_1$; this hints to the existence of a symmetry in the potential, i.e., a non-physical phase. One can rotate this unphysical phase away, and define the state $\eta'_1 = \eta_2 - \eta_1$. Such a transformation is equivalent to taking the $\eta_2 \to 0$ limit, and we assume this limit to be taken when writing the fields in terms of components in Eq.~\eqref{eq:explicit-fields}.

To facilitate the analysis, we apply a conformal transformation from the Jordan frame, which contains terms with scalar-gravity couplings, to the Einstein frame with no explicit couplings to gravity \cite{Kaiser:2010ps}.  Physical observables are invariant under this frame transformation. The two frames are equivalent after the end of inflation when the transformation parameter equals unity.

\subsubsection*{Going from the Jordan frame to the Einstein frame}

In terms of of fields $h_1$, $\eta_1$, and $h_2$, the action in the Jordan frame can be written as 
\bea
\label{eq:action-Jordan-2}
S_J &=& \int d^4x \sqrt{-g}
\bigg[  \frac{1}{2} \biggl( \MPl^2 + \xi_1 \left( h_1^2 +\eta_1^2 \right) + \xi_2 h_2^2 + 2|\xi_4| \left( h_1h_2 c_{\theta_4}+\eta_1h_2 s_{\theta_4} \right) \biggr) R \nonumber\\
&&\hspace{22mm} - \frac{1}{2}g^{\mu\nu}\biggl(\partial_\mu h_1  \partial_\nu h_1
+ \partial_\mu \eta_1  \partial_\nu \eta_1 + \partial_\mu h_2  \partial_\nu h_2 \biggr)
- V(h_1, \eta_1,h_2) \biggr]\,,
\eea
where the last term is the potential in Eq.~\eqref{eq:approxscalarpot}, which could be simplified; since the quadratic terms are subdominant at high field values we keep only terms that are quartic in $h_{1,2}$ and $\eta_{1}$.  This reduces the potential to
\bea
V(h_1,\eta_1,h_2)  &\approx & \frac{1}{4} \biggl[
\lambda_{11} (h_1^2 + \eta_1^2)^2
+\lambda_{22}  h_2^4
+(\lambda_{12}+\lambda'_{12})(h_1^2 + \eta_1^2)h_2^2
\nonumber\\
&& ~~~~~~
+ 2 |\lambda_1| \biggl( c_{\theta_1}
\left(h_2^2(h_1^2 - \eta_1^2) \right)
+ 2\, s_{\theta_1} h_2^2h_1\eta_1  \biggr)
 \biggr]\,,
\label{eq:def-V}
\eea
where $\theta_1$ is the phase of the $\lambda_1$ parameter, and $c_{\theta_k} = \cos\theta_k$ and $s_{\theta_k} = \sin\theta_k$. In order to simplify our analysis and keep the focus on the primordial CP violation (rather than dealing with the intricacies of a multi-field inflationary scenario), we consider a proportional solution in which the following relations hold
\be
\label{eq:B1B2}
\eta_1 = \beta_1 \, h_1\,,
\qquad
h_2 = \beta_2 \, h_1 \, ,
\ee
where $\beta_1,\beta_2$ are dependant on the angles $\theta_1,\theta_4$ and the $\lambda_i$ parameters of the potential (as will be shown later in Eqs.~(\ref{eq:B2atXmin}-\ref{eq:B1atXmin}). In this limit, the model is reduced to the familiar single-field inflation model and the potential can be written simply as
\be 
V(h_1)= \frac{\widetilde{\lambda}}{4}\, h_1^4 
\label{eq:def-V-simp}
\ee
where
\be
\label{eq:def-tilde-lambda}
\widetilde{\lambda} =
 \lambda_{11}  (1+ \beta_1^2)^2 +\lambda_{22} \,  \beta_2^4  + \biggl(
(\lambda_{12}+\lambda'_{12})(1 + \beta_1^2)
+ 2 |\lambda_1| \left( c_{\theta_1}
(1 - \beta_1^2) + 2\, s_{\theta_1} \beta_1 \right)\biggr) \beta_2^2 \,.~~
\ee

In order to have the action in the Jordan frame resemble the action for 1HDM-inflation, we make some additional rescalings:
\bea \label{eq:def-tilde-xi}
\widetilde{\xi} & = & \xi_1(1+\beta_1^2)+\xi_2\beta_2^2+2 \,|\xi_4|
\beta_2(c_{\theta_4} +\beta_1 s_{\theta_4})  \nonumber \\
\widetilde{\zeta} & = & 1+\beta_1^2+\beta_2^2 \nonumber \\
\widetilde{r} & = & \widetilde{\xi}/\widetilde{\zeta} \nonumber \\
\widetilde{h}_1 & = & \sqrt{\widetilde{\zeta}}\ h_1 \ .
\eea
These rescalings render the Jordan-frame action as
\be
S_J = \int d^4x \sqrt{-g}
\left[   \frac{\MPl^2}{2} R \left(1+\widetilde{r}\frac{\widetilde{h}_1^2}{\MPl^2}\right) - \frac{1}{2}g^{\mu\nu}\partial_\mu \widetilde{h}_1  \partial_\nu \widetilde{h}_1  - \frac{\widetilde{\lambda}}{4} \frac{\widetilde{h}_1^4}{\widetilde{\zeta}^2} \right]\,.
\label{eq:S_J}
\ee
(We note that $\widetilde{r}\ \widetilde{h}_1^2=\widetilde{\xi}\ h_1^2$.)

After a conformal transformation to the Einstein frame, $g_{\mu\nu}\to \Omega^{-2}g_{\mu\nu}$ where $\Omega^2 = 1+ \widetilde{r}\ \widetilde{h}_1^2/\MPl^2$, and in the limit $6\,\widetilde{r}\gg 1$, we can define a dimensionless scalar field $\At$ as
\be
\At=\sqrt{6}\, \ln\sqrt{1+\widetilde{r}\ \frac{\widetilde{h}_1^2}{\MPl^2}} \,,
\ee
and the Einstein frame action is (here, $R$ and $g^{\mu\nu}$ are the Ricci scalar and metric tensor in the Einstein frame) 
\be
\label{eq:S_E}
S_E = \int d^4x \sqrt{-g}
\left[   \frac{\MPl^2}{2} R - \frac{1}{2}g^{\mu\nu}\MPl^2\,\partial_\mu \At \,  \partial_\nu \At  - \frac{\widetilde{\lambda}}{4} \frac{\MPl^4}{\widetilde{\xi}^2} \left(1-e^{-2\At/\sqrt{6}}\right)^2\right]\,.
\ee
The potential is the Einstein frame,
\be
\frac{\widetilde{\lambda}}{4} \frac{\MPl^4}{\widetilde{\xi}^2} \left(1-e^{-2\At/\sqrt{6}}\right)^2 \equiv \Vt({\At}) \,,
\ee
is a result familiar from 1HDM inflation. 

We will be interested in the effect of the non-minimal coupling $\xi_4$ and the associated phase $\theta_4$. Therefore, we will set $\xi_1=\xi_2=0$ and assume that the initial field values are such that $\Omega^2>0$ is guaranteed.  Therefore, with these assumptions, the parameter $\xit$ from Eq.~\eqref{eq:def-tilde-xi} is reduced to
\be
\xit= 2 \,|\xi_4| \,\beta_2(c_{\theta_4} +\beta_1 s_{\theta_4}) \,,
\ee 
and the potential $\Vt({\At})$ in Eq.~\eqref{eq:S_E} takes the form 
\be
\label{eq:full-pot-B1B2}
\Vt(\At) =
\biggl(\frac{\MPl^2}{2 \,|\xi_4|} \biggr)^2
X(\beta_1,\beta_2)\,
\left(1-e^{-2\At/\sqrt{6}}\right)^2 ,
\ee
where we have collected all $\beta_1,\beta_2$ proportionality in the function $X(\beta_1,\beta_2)$ as
\be
X(\beta_1,\beta_2) =
\frac{\widetilde{\lambda} }
{4\,\beta_2^2(c_{\theta_4} + \beta_1\, s_{\theta_4})^2 } \,.
\ee

To find the minimum of the inflationary potential, we use the procedure in Ref.~\cite{Gong:2012ri} for a 2HDM. We first minimize  $X(\beta_1,\beta_2)$ with respect to $\beta_2$, which occurs at $\partial X(\beta_1,\beta_2)/\partial \beta_2=0$, yielding
\be
\beta_2^2 = \sqrt{\frac{\lambda_{11}}{\lambda_{22}} } \, \left(1+\beta_1^2\right) \,.
\label{eq:B2atXmin}
\ee
We repeat the same treatment and minimise the $X(\beta_1)$ function with respect to $\beta_1$, which results in:
\bea
\label{eq:B1atXmin}
\beta_1 & = &
\frac{(\Lambda +2 |\lambda_1| c_{\theta_1}) s_{\theta_4} - 2 |\lambda_1| c_{\theta_4} s_{\theta_1}}{(\Lambda -2 |\lambda_1| c_{\theta_1}) c_{\theta_4} - 2 |\lambda_1| s_{\theta_4} s_{\theta_1}}\,,
\eea
with $\Lambda = \lambda_{12}+\lambda'_{12}+2\sqrt{\lambda_{11}\lambda_{22}}\,$.  Using the $\beta_2$ value in Eq.~\eqref{eq:B2atXmin}, we can write the $X(\beta_1,\beta_2)$ function solely in terms of $\beta_1$:
\be
X(\beta_1)=
\frac{
(1+ \beta_1^2) \, \Lambda + 2  \left[ (1-\beta_1^2) c_{\theta_1} + 2\beta_1 s_{\theta_1}\right] |\lambda_1|}{4\,(c_{\theta_4} + \beta_1\, s_{\theta_4})^2 } \,.
\ee
Replacing the $\beta_1$ value which minimizes $X(\beta_1,\beta_2)$ back into the $X(\beta_1)$ function itself, yields the form of $X$ independent of $\beta_1$ and $\beta_2$ with only $\theta_1$ and $\theta_4$ as variables:
\be
\label{eq:X-thetas}
X(\theta_1,\theta_4) = \frac{\frac{1}{4}\Lambda^2 - \lambda_1^2  }{\Lambda - 2\lambda_1 \cos(\theta_1-2\theta_4)}
\ee
The function $X(\theta_1,\theta_4)$ depends on two angles (we take the ranges $-\pi<\theta_1<\pi$ and $0<\theta_4<\pi$) and five quartic couplings ($\lambda_1,\ \lambda_{12},\ \lambda'_{12},\ \lambda_{11}$, and $\lambda_{22})$.  With $\beta_1$ and $\beta_2$ fixed (hence $\eta_1$ and $h_2$ fixed in terms of $h_1$), the potential is now a function of a single field $\At$.  Finally, we arrive at
\be
\Vt(\At) = \biggl(\frac{\MPl^2}{2 \,|\xi_4|} \biggr)^2 \left(1-e^{-2\At/\sqrt6}\right)^2 X(\theta_1,\theta_4) \,.
\label{eq:VofA}
\ee

In the next section, we show that agreement with CMB measurements (for example Planck 2018~\cite{Planck:2018jri}) requires $X(\theta_1,\theta_4)/|\xi_4|^2 \sim 3.8\times10^{-10}$.   Since seven parameters enter $X(\theta_1,\theta_4)$ (along with the unknown $\xi_4$), it is not difficult to engineer this result.  For example, if we assume that all of the quartic couplings are of the same order, say $\lambda_\mathrm{3HDM}$, then roughly $X(\theta_1,\theta_4)\sim \lambda_\mathrm{3HDM}$, and we need $|\xi_4| \sim 5\times10^4\sqrt{\lambda_\mathrm{3HDM}}$.\footnote{Stability of the potential requires $\lambda_{ii} > 0, \ \lambda_{ij} + \lambda'_{ij} > -2 \sqrt{\lambda_{ii}\lambda_{jj}}, \ |\lambda_i| \leq |\lambda_{ii}|, |\lambda_{ij}|, |\lambda'_{ij}| , \ i\neq j = 1,2,3$.  It is also reasonable to require the quartic couplings to be less than unity.}

In the 1HDM inflation model, there is one quartic coupling, $\lambda_\mathrm{1HDM}$, and one nonminimal coupling parameter, $\xi_\mathrm{1HDM}$, and the corresponding requirement is $\xi_\mathrm{1HDM} \sim 5\times10^4\sqrt{\lambda_\mathrm{1HDM}}$. The 1HDM model constrains $\lambda_\mathrm{1HDM}=0.12$, which demands and $\xi_\mathrm{1HDM}\approx 10^4$.  Such a large value of $\xi_\mathrm{1HDM}$ has issues related to unitarity \cite{Barbon:2009ya,Burgess:2010zq,Hertzberg:2010dc}.  But in the 3HDM inflation setup the quartic couplings associated with the inert doublets can be much smaller than 0.12, and a much smaller value of $\xi_4$ can be realized, evading, or at least mitigating, the unitarity issue.

\begin{figure}[!ht]
\centering
\includegraphics[width=0.49\linewidth]{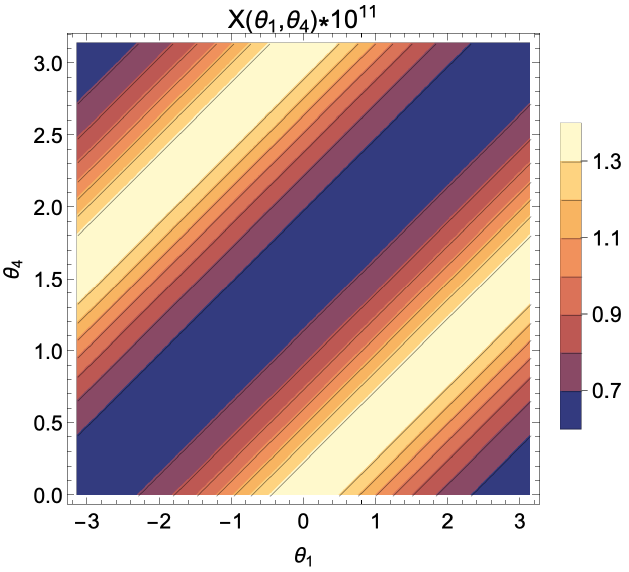}
\caption{The values of the $X(\theta_1,\theta_4)$ function (multiplied by $10^{11}$) for varying values of $\theta_1$ and $\theta_4$ using the values of the quartic couplings in the text.}
\label{Fig:X-T1T4-B1B2}
\end{figure}

For an explicit example of the choice of the quartic couplings satisfying vacuum stability, consider the following illustrative choices of parameters: $\{|\xi_4|, \lambda_1, \lambda_{12}, \lambda'_{12}, \lambda_{11}, \lambda_{22}, \theta_1,\theta_4\} = \{ 1/6, 8\times10^{-13},9\times10^{-12},1\times10^{-11},1.1\times10^{-11},1.15\times10^{-11},\pi/3,\pi/3\}$ which yields $X(\theta_1,\theta_4)=10^{-11}$. Fig.~\ref{Fig:X-T1T4-B1B2} shows the $X(\theta_1,\theta_4)$ function for allowed values of $\theta_1$ and $\theta_4$ using the values of the quartic couplings above as an example.  For $|\xi_4|=1/6$, the desired value is $X(\theta_1,\theta_4) \sim 10^{-11}$.  This is the value displayed in Fig.~\ref{Fig:X-T1T4-B1B2}.

\section{Inflationary dynamics for the 3HDM}
\label{sec:slow-roll}

\begin{figure}[!ht]
\centering
\includegraphics[height=4.8cm]{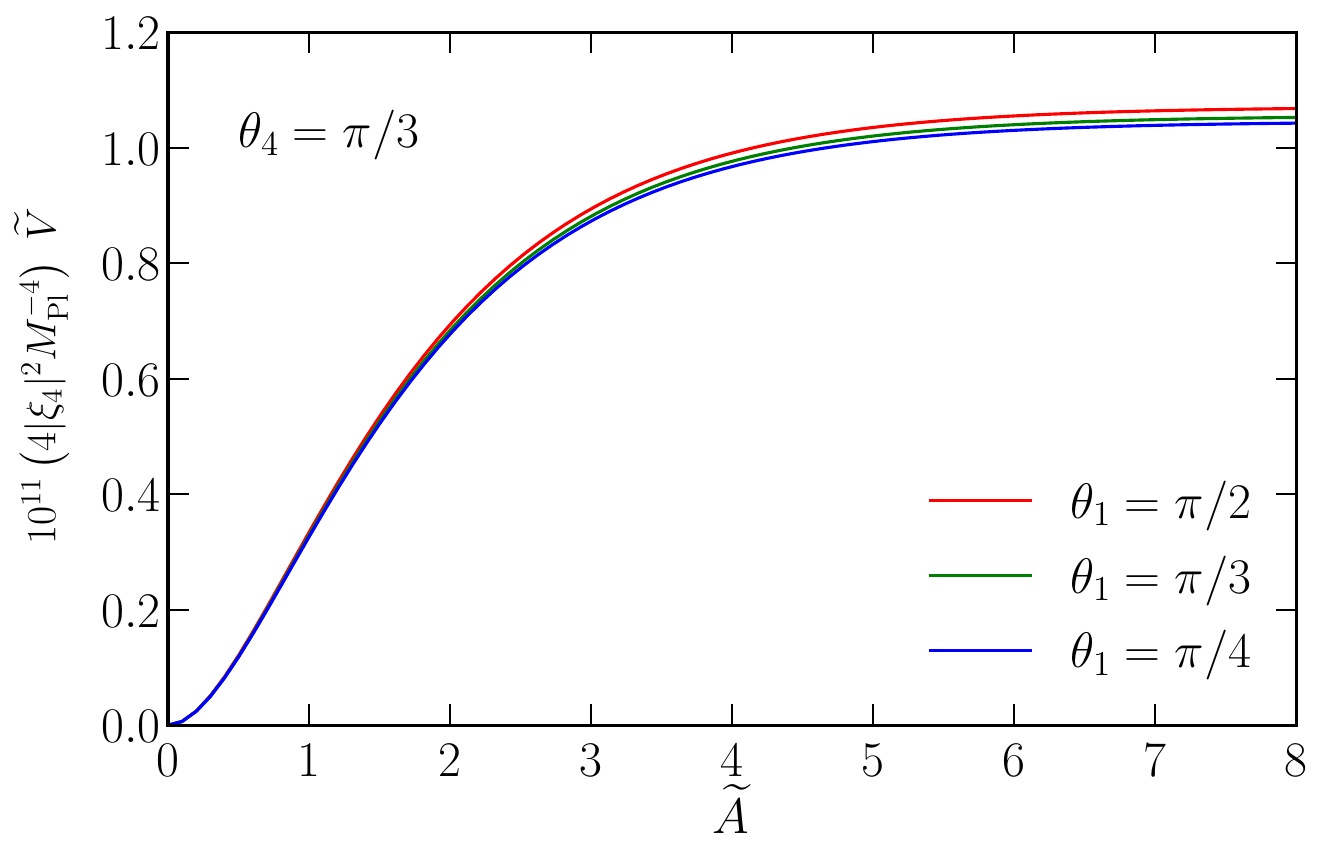} 
\includegraphics[height=4.8cm]{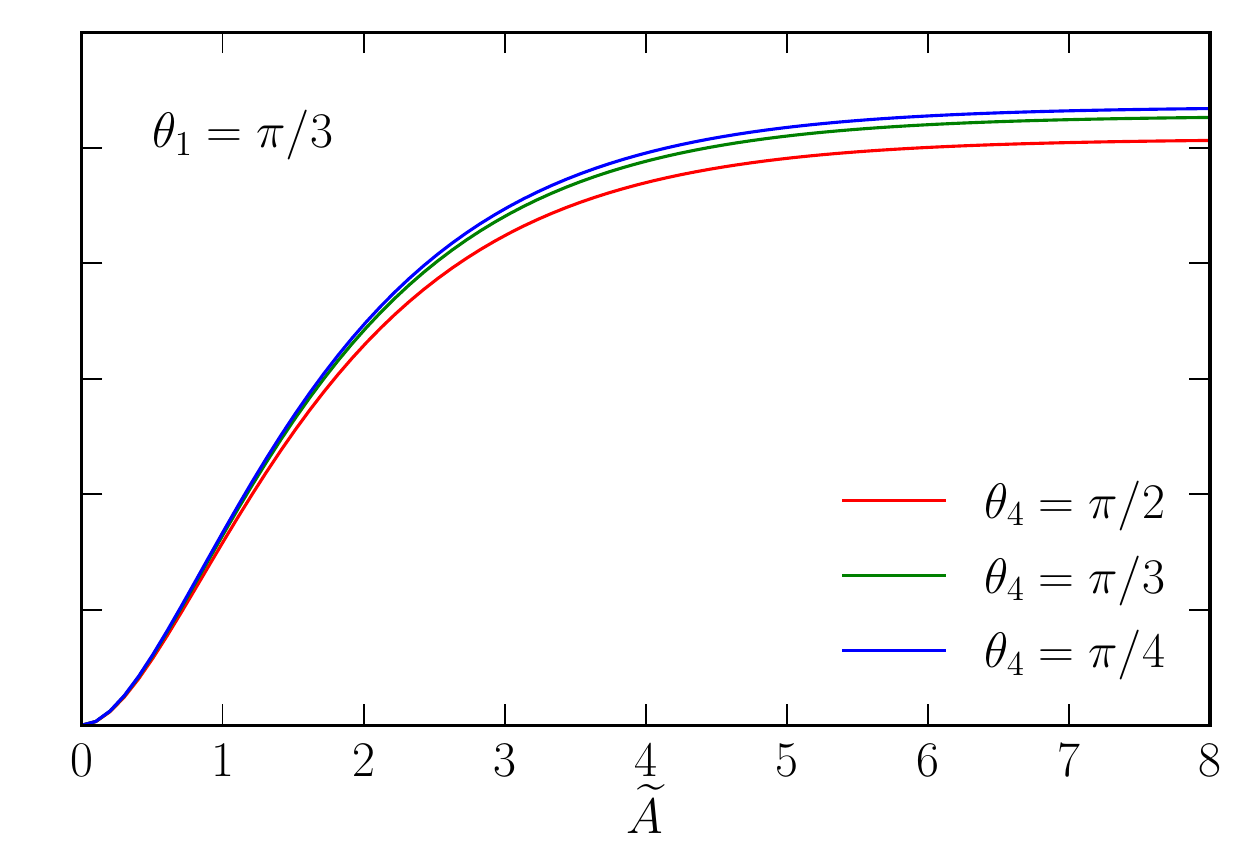}
\caption{The inflationary potential as a function of $\At=A/\MPl$ for different values of $\theta_1$ and $\theta_4$ (using the quartic couplings listed in the text).}
\label{Fig:V-vs-ts}
\end{figure}

Figure \ref{Fig:V-vs-ts} shows the inflationary potential, given in Eq.~\eqref{eq:VofA}, for different values of  $\theta_1$ and $\theta_4$. Note that the potential is almost flat at high field values, which ensures a slow roll inflation. The figure also demonstrates that the potential is not very sensitive to $\theta_1$ and $\theta_4$.

For the calculation of the slow roll parameters the function $X(\theta_1,\theta_4)$ and the factor $\MPl^4/4|\xi_4|^2$ are irrelevant since they cancel in the expressions for $\epsilon_V$ and $\eta_V$, which are
\bea
\epsilon_V &=&
\frac{1}{2}\MPl^2 \,\left(\frac{1}{\Vt}\frac{d\Vt}{dA}\right)^2 = 
\frac{4}{3}\frac{1}{\left( 1- e^{2\At/\sqrt{6} }  \right)^2}\,,
\label{eq:epsilon}
\\[3mm]
\eta_V &=&
\MPl^2 \, \frac{1}{\Vt}\left(\frac{d^2\Vt}{dA^2} \right) = 
\frac{4}{3} \frac{2-e^{2\At/\sqrt{6}}} {\left(1- e^{2\At/\sqrt{6}}\right)^2} \, ,
\label{eq:eta}
\eea
where $A=\MPl \, \At$.  For field values of $\At\gg 1 $ (or equivalently, $A\gg \MPl$), both parameters $\epsilon_V,\eta_V \ll 1$, which satisfies the slow roll condition. Inflation ends when $\epsilon_V=1$, or when $\At= \At_e = 0.94$.

In terms of $\Vt$ and the slow-roll parameters, the amplitude of the scalar power spectrum, $A_s$, the tensor to scalar ratio, $r$, and the scalar spectral index, $n_s$ are given by 
\bea
A_s & = & \frac{1}{24\pi^2} \frac{1}{\epsilon_V} \frac{\Vt}{\MPl^4} 
= \frac{1}{128\pi^2}\frac{(1-e^{2\At/\sqrt{6}})^4}{e^{4\At/\sqrt{6}}}\frac{X(\theta_1,\theta_4)}{|\xi_4|^2} \,,  
\nn \\
r   & = & 16 \epsilon_V \,, \nn \\
n_s & = & 1-6\epsilon_V+2\eta_V \,. \label{eq:A_r_n}
\eea

\begin{figure}[!ht]
\centering
\includegraphics[width=0.7\linewidth]{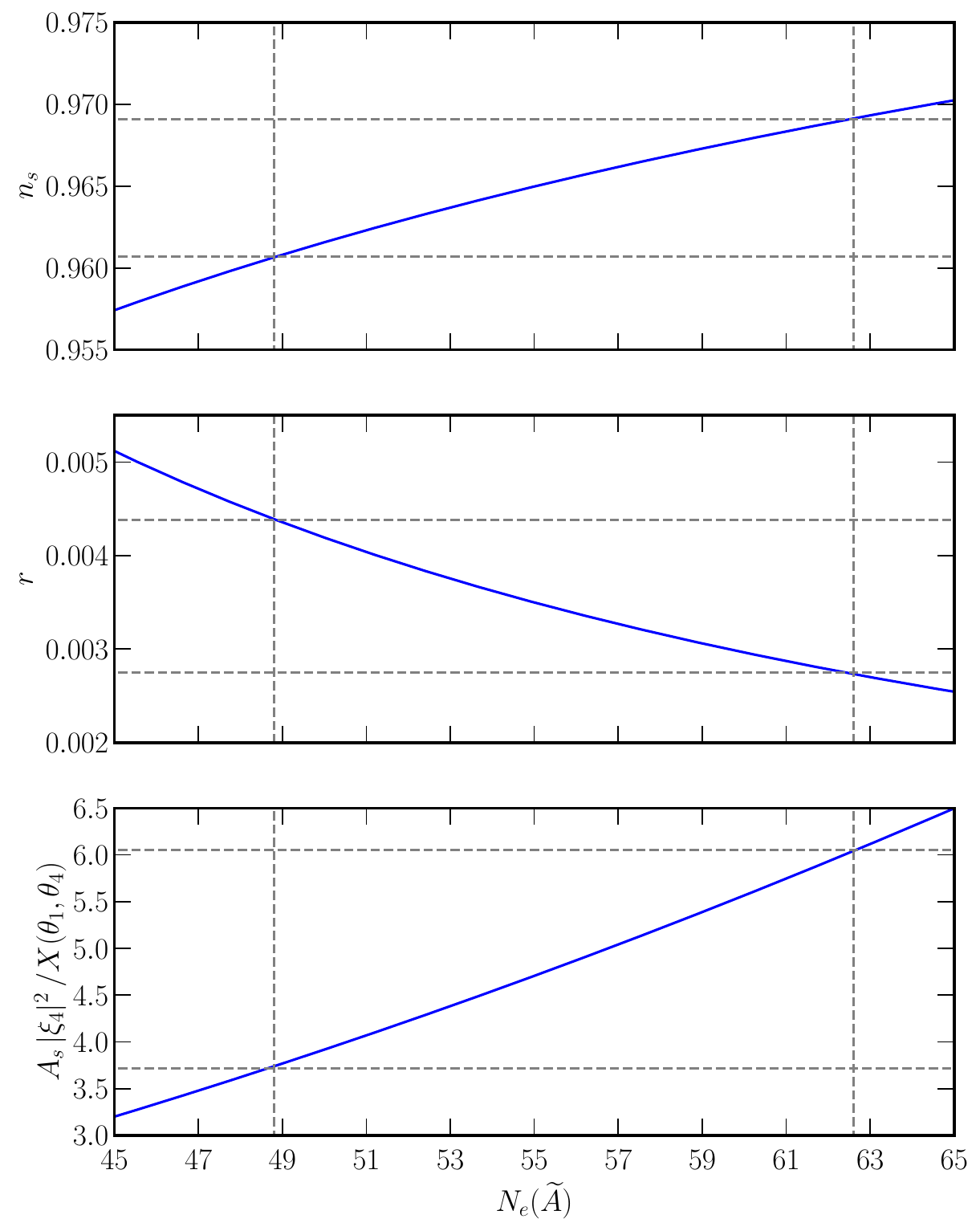}
\caption{The spectral index $n_s$ (top), the tensor-to-scalar ratio $r$ (middle),  and the scalar amplitude $A_s$ (bottom) as a function of $N_e(\At)$.  The horizontal dashed lines for $n_s$ represent the Planck-18 limits $n_s=0.9649\pm0.0042$, and the vertical dashed lines are the corresponding values of $N_e(\At)$.  The regions of $N_e(\At)$ consistent with the limits on $n_s$ are also shown in the middle and bottom figures.  The corresponding limit on $r$ satisfies the Planck bound ($r<0.06$).  The bottom figure illustrates that $3.7<A_s|\xi_4|^2/X(\theta_1,\theta_4) < 6.1$.}
\label{Fig:Ps-r-ns}
\end{figure}

Next we wish to calculate as a function of $\At$ the number of $e$-folds of expansion before the end of inflation:
\be
\label{eq:efolds}
N_e(\At) = \frac{1}{\MPl^2}\int_{A_e}^{A} \frac{\Vt}{\Vt'}\,dA
=
\frac{3}{4} \left[ \frac{2}{\sqrt{6}}\At_e - \frac{2}{\sqrt{6}}\At - e^{2\At_e/\sqrt{6}} + e^{2\At/\sqrt{6}} \right] \,.
\ee
For a given value of $\At$, we can use Eq.~\eqref{eq:efolds} to calculate $N_e(\At)$, Eq.~\eqref{eq:epsilon} and Eq.~\eqref{eq:eta} to calculate $\epsilon_V$ and $\eta_V$, and Eq.~\eqref{eq:A_r_n} to calculate $n_s$, $r$, and $A_s$.  The results for $n_s$, $r$, and $A_s$ as a function of $N_e(\At)$ are given in Fig.~\ref{Fig:Ps-r-ns}.

We can use the limits on $A_s$ to get an idea of the required value of $X(\theta_1,\theta_4)/|\xi_4|^2$.  The Planck 2018 limit on $A_s$ is $\ln(A_s)=3.044\pm0.014$~\cite{Planck:2018jri}.  For an estimate, we take the central vale of $A_s=2.1\times10^{-9}$, and choose $A_s|\xi|^2/X(\theta_1,\theta_4)=5.6$ from the allowed region of Fig.~\ref{Fig:Ps-r-ns}.  Thus, agreement with CMB data would be satisfied for 
\be
\frac{X(\theta_1,\theta_4)}{|\xi_4|^2} \sim 3.8\times10^{-10} \,.
\ee
This was the value used in Sec.~\ref{sec:Ip3HDM}, for which $|\xi_4|=1/6$.

At this point we can also check the field values in terms of the original field $h_1$.  Using the value of $\At$ at the end of inflation, $\At_e=0.94$, we find
\be
\label{eq:hi&hf}
\frac{{h_1}(\At_e)}{\MPl}  =
\frac{0.82}{\sqrt{|\xi_4|\,\beta_2(c_{\theta_4} + \beta_1\, s_{\theta_4}) }}\,.
\ee

To conclude this section we note that the 3HDM inflation model with two inert doublets and the transformations under the $Z_2$ group of $g_{Z_2}=  \mathrm{\rm diag}\left(-1, -1, +1 \right)$ is effectively a single-field inflation model with a potential given by Eq.~\eqref{eq:full-pot-B1B2}.  This model is consistent with CMB observations for the choice $X(\theta_1,\theta_4)/|\xi_4|^2 \sim 3.8\times10^{-10}$.  The function $X(\theta_1,\theta_4)$ involves five quartic couplings and two angles.  Assuming the quartic couplings are of the same order of magnitude, the requisite value of $X(\theta_1,\theta_4)/|\xi_4|^2$ may be realized by choosing the quartic couplings small, say $\lambda=\mathcal{O}(10^{-11})$, and the nonminimal coupling $|\xi_4|=\mathcal{O}(1/6)$, or having the quartic couplings the same order as in the 1HDM, $\lambda=0.12$, with the nonminimal coupling $|\xi_4|=\mathcal{O}(10^4)$.  The advantage of the 3HDM inflation model over the 1HDM inflation model is that it can avoid large nonminimal couplings and the concomitant unitarity issue (provided the quartic couplings are very small).

But the real advantage of the 3HDM inflation model is the appearance of a complex quartic coupling $\lambda_1$, which will allow for baryogenesis after inflation\footnote{Note that all four parameters $\mu^2_{12}, \, \lambda_1, \, \lambda_2$ and $\lambda_3$ can be complex, however, $\lambda_1$ is the only one affecting the inflationary potential. The complex phase of other parameters affect the reheating process, which we discuss in an upcoming publication.}. Moreover, the 3HDM is the most minimal model allowing for a complex non-minimal coupling to gravity, $\xi_4$ which also contributes to baryogenesis.

\section{Baryogenesis from scalar asymmetries}
\label{sec:baryogenesis}

\subsection{Scalar asymmetries}

After inflation the energy density of the Universe is stored in the inflaton fields $\phi_1,\phi_2$.  Through the reheating process, this energy density is eventually transferred to the SM fields.  The first step in the process is conversion of $\phi_1,\phi_2$ to the active doublet $\phi$ through the potential terms given in Eq.~\eqref{eq:VofA} which are responsible for producing a  CP asymmetry in $\phi$, which is then transferred to an asymmetry in the SM degrees of freedom.

To discuss the consequences of complex phases we assume instant reheating. Since the field $\phi$ is light with respect to the inflaton degrees of freedom, we expect the latter to quickly annihilate to $\phi$. The end product will be the creation of an unequal number of $\phi$ and $\phi^\ast$ quanta as follows.

\begin{figure}[!ht]
\centering
\includegraphics[height=1.5in]{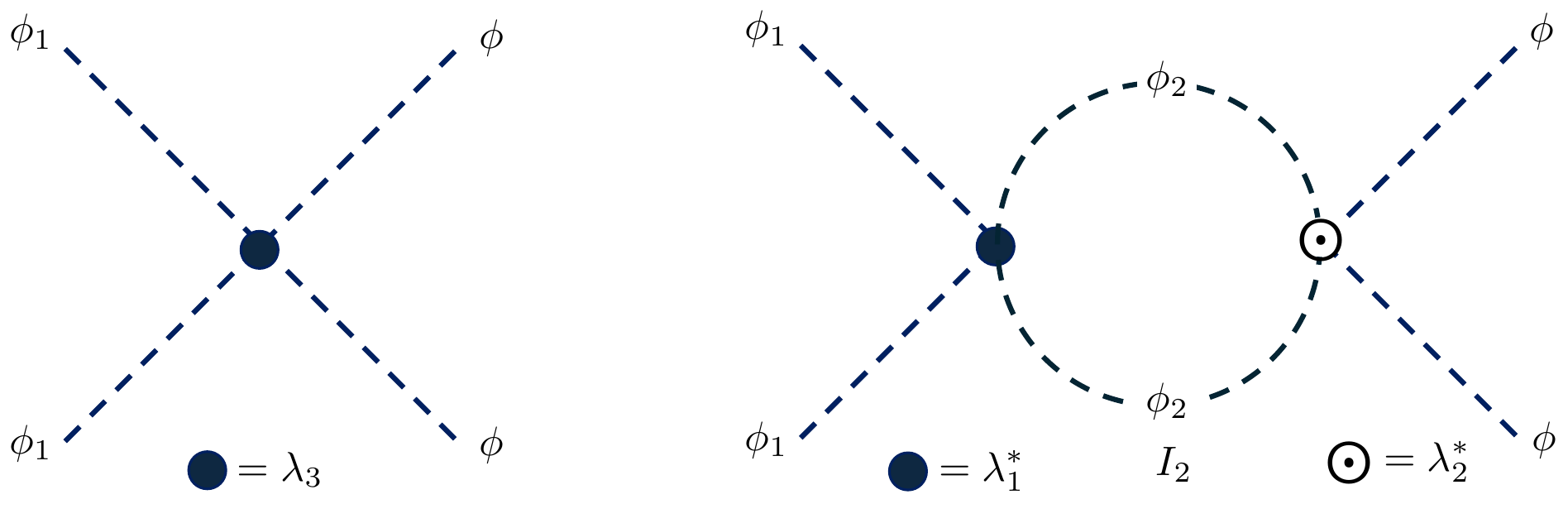}
\caption{On the left is the tree-level process $\phi_1\phi_1 \to \phi \phi$, and on the right, the one-loop bubble diagram with $\phi_2$ in the loop, yielding an absorptive part to the amplitude since the particles in the loop can propagate on shell \cite{Yanagida:1979fb, Nanopoulos:1979gx, Harvey:1981yk, Kolb:1983ni, Balaji:2004xy, Balaji:2005ha}. }
\label{fig:tree-loop} 
\end{figure}

To show how the CP-violating couplings of the inflaton lead to a scalar asymmetry, we focus on the annihilation process $\phi_1 \phi_1 \to \phi \phi$ and the complex conjugated process.\footnote{This is similar in spirit to the approach in Refs.\ \cite{Balaji:2004xy,Balaji:2005ha}.}  From the potential in Eq.~\eqref{eq:V0-3HDM}, one can calculate the amplitude of these tree-level processes to be 
\be
\mathcal{M}_{\phi_1 \phi_1 \to \phi \phi}^\mathrm{tree} \; \propto \; \lambda_3 \, 
\qquad \mbox{and} \qquad 
\mathcal{M}_{\phi_1^*\phi_1^* \to \phi^* \phi^*}^\mathrm{tree} \; \propto \;  \lambda_3^*  \,.
\ee
The generation of the asymmetry is sensitive to the interference between the tree and loop diagrams.  At one loop level, there are many diagrams that contribute to this process. For the purpose of demonstration, we consider the bubble diagrams which convert $\phi_1\phi_1$ to $\phi\phi$ with only $\phi_2$ in the loop, as shown in Fig.~\ref{fig:tree-loop}. Clearly, one needs to take into account all diagrams contributing to this decay process, specially since there may be interferences cancelling the CP asymmetry. However, since all scalar couplings in the potential can be different, one can ensure that such cancellation does not occur.  More careful analysis of these effects is deferred to a future work. 

The amplitude of the loop process with $\phi_2$ running in the loop is proportional to 
\be
\mathcal{M}_{\phi_1 \phi_1 \to \phi \phi}^\mathrm{loop} \propto \lambda_1^*\,I_2\, \lambda_2^*  \, 
\qquad \mbox{and} \qquad 
\mathcal{M}_{\phi_1^* \phi_1^* \to \phi^* \phi^*}^\mathrm{loop} \propto \lambda_1\,I_2\, \lambda_2  \,.
\ee

Due to the interference of the tree and loop diagrams, the decay processes are CP violating and result in unequal number of $\phi$ and $\phi^*$ states. We define the asymmetry $A^1_{CP}$ as the difference between the $\phi_1$ decay rate and its conjugate, and we find 
\bea
A^1_{CP} &\propto& \left|\mathcal{M}_{\phi_1 \phi_1 \to \phi \phi}^\mathrm{tree} + \mathcal{M}_{\phi_1 \phi_1 \to \phi \phi}^\mathrm{loop}   \right|^2 - 
\left|\mathcal{M}_{\phi_1^* \phi_1^* \to \phi^* \phi^*}^\mathrm{tree} + \mathcal{M}_{\phi_1^* \phi_1^* \to \phi^* \phi^*}^\mathrm{loop}   \right|^2 \nonumber \\[1ex]
&\propto& \mathrm{Im}\left[\lambda_1\lambda_2\lambda_3\right] \, \mathrm{Im}[I_2]
\propto - |\lambda_1| \, |\lambda_2| \, |\lambda_3| \, \sin(\theta_1+\theta_2+\theta_3)\, \mathrm{Im}[I_2] \ .
\label{eq:ACP}
\eea
This result illustrates that in order to produce an asymmetry in $\phi$, two conditions must be met: 1) $\mathrm{Im}\left[\lambda_1\lambda_2\lambda_3\right]\neq0$, or equivalently $\theta_1+\theta_2+\theta_3 \neq n\pi$, where $n$ is an integer;  2) $\mathrm{Im}[I_2] \neq 0$.  The last condition is satisfied if $m_{\phi_2} < m_{\phi_1}$.  (If, instead, $m_{\phi_2} > m_{\phi_1}$, an expression similar to Eq.~(\ref{eq:ACP}) would obtain mutatis mutandis.) 

This asymmetry in the scalar sector will be then transferred to the fermion sector as described in detail in Sec.~\ref{sec:baryogenesis2}.

\subsection{Baryogenesis}
\label{sec:baryogenesis2}
In this subsection we demonstrate how the asymmetry produced in the number densities of $\phi$ and $\phi^*$ can translate into an asymmetry in the baryon number.  Our analysis closely follows the original analysis of Ref.~\cite{Harvey:1990qw}, or more recently, follows Refs.~\cite{Servant:2013uwa} and \cite{Davidson:2013psa}. 

The asymmetry in particle species $i$, denoted as $N_i$, can be written in terms of the number densities of particles and antiparticles ($n_i$ and $n_{\overline{i}}$) as
\be 
\label{eq:chem-pot}
N_i \equiv n_i - n_{\overline{i}} = \frac{g_i T^2}{6} \, \mu_i \times \left\{
\begin{array}{cl}
2 & \textrm{for bosons}\\
1 & \textrm{for fermions}
\end{array}
\right.,
\ee
where $T$ is the temperature, $g_i$ is the number of degrees of freedom of a particle, and $\mu_i$ is its chemical potential.  Assuming that the primordial thermal bath contains SM fermions, SM gauge bosons, and one Higgs doublet, $\phi$, we consider temperatures after reheating (well before the electroweak phase transition) when all the Yukawa interactions are in equilibrium. Moreover, the gauge symmetries at these temperatures are unbroken, leading to zero chemical potential for the gauge bosons. To show that the asymmetry produced in $\phi$ versus $\phi^*$ can be used to generate a baryon asymmetry, let us calculate the chemical potential for all particles in the primordial thermal bath.

Note that when an interaction is in chemical equilibrium, the sum of the chemical potentials of all particles involved vanishes. Chemical potentials for the fields are defined to be
\be
\mu_{W^-} = \mu_W,\ \ \mu_{\phi^0_3} = \mu_0,\ \ \mu_{\phi^-_3} = \mu_-, \ \  \mu_{u_L}, \ \ \mu_{d_L}, \ \ \mu_{e_L}, \ \  \mu_{\nu_L}, \ \  \mu_{u_R}, \ \ \mu_{d_R}, \ \mu_{e_R} \, .
\ee
With interactions in equilibrium, the following terms lead to the indicated relations among the chemical potentials: 
\bea
W^-\,\phi^+\phi^0_3 ~~&\Rightarrow &~~\mu_W -\mu_- -\mu_0 =0 , \label{eq:Wphiphi} \\
W^-\,u_L\,\bar{d}_L ~~&\Rightarrow &~~\mu_W + \mu_{u_L} - \mu_{d_L} =0 , \label{eq:Wud} \\
W^-\,\nu_L\,\bar{e}_L ~~&\Rightarrow &~~\mu_W + \mu_{\nu_{L}} - \mu_{e_L} =0 , \label{eq:Wnue} \\
\bar{Q}_L\cdot\phi\ d_R  ~~&\Rightarrow &~~ - \mu_{d_L} + \mu_0 + \mu_{d_R} = 0, \label{eq:yuk_d}
\\
\epsilon^{ab}\,\bar{Q}_{La}\ \Phi_{3b}^\dagger\ u_R ~~&\Rightarrow &~~ - \mu_{u_L} - \mu_0 + \mu_{u_R} = 0,
\label{eq:yuk_u}
\\
{\bar{\ell}}_L\cdot\phi\, e_R ~~&\Rightarrow &~~ - \mu_{e_L} + \mu_0 + \mu_{e_R} = 0,
\label{eq:yuk_e}
\eea
where $Q_L$ ($\ell_L$) represents the left-handed quark (lepton) doublet, $d_R$ ($u_R$) the right-handed down-type (up-type) quark singlet, and $e_R$ the right-handed lepton singlet fields.  It is understood that Eqs.~(\ref{eq:Wud}-\ref{eq:yuk_e}) apply for each generation of quarks and leptons.  We will assume there are rapid flavor-changing interactions, so the chemical potentials for all generations of quark and lepton fields are equal.\footnote{We now know that neutrinos have mass and lepton flavor is not conserved, so we will also assume equal chemical potentials for each neutrino flavor.}   

Denoting the number of generations as $N_g=3$ and including factors of $3$ for quark colors, the charge density is calculated to be 
\bea 
Q &=& N_g \left( Q_{e}N_{e_R}  +Q_{e} N_{e_L} + 3Q_{u}N_{u_L} + 3Q_{u}N_{u_R} +3Q_{d}N_{d_L} + 3Q_{d}N_{d_R} \right) \nonumber \\
& & + Q_{\phi_{3^-}} N_{\phi^-} +Q_{W^-}N_{W^-} \,, \nonumber\\
Q&\propto& - 3\mu_{e_R}  - 3\mu_{e_L} + 6\mu_{u_L} + 6\mu_{u_R} -3\mu_{{d_L}} -3\mu_{d_R} -2\mu_-  -4\mu_W \,, \nonumber\\
Q&\propto& 6\mu_{u_L} - 6\mu_{\nu_L} - 18\mu_W + 14\mu_0 \, , 
\eea 
where we have used the relations in Eqs.~(\ref{eq:Wphiphi}-\ref{eq:yuk_e}). 

The density of the third component of weak isospin is
\bea
\mathcal{T}^3 &=& N_g \left( \mathcal{T}^3_{e_L}N_{e_L} + \mathcal{T}^3_{\nu_L}N_{\nu_L} \, + 3\mathcal{T}^3_{u_L}N_{u_L} + 3\mathcal{T}^3_{d_L}N_{d_L} \right) + \mathcal{T}^3_{\phi_{3^0}} N_{\phi^0} + \mathcal{T}^3_{\phi_{3^-}}N_{\phi^-} + \mathcal{T}^3_{W^-} N_{W^-} \,, \nonumber \\
\mathcal{T}^3 &\propto&  - \frac{3}{2}\mu_{e_L} + \frac{3}{2}\mu_{\nu_L} + \frac{9}{2}\mu_{u_L} - \frac{9}{2}\mu_{d_L} - \mu_0 - \mu_- -4\mu_W \,, 
\nonumber \\ [1ex]
\mathcal{T}^3 &\propto& -11\mu_W \, . \label{eq:T3} 
\eea

Let us now turn to the baryon and lepton number comoving asymmetries. The comoving baryon asymmetry, $\mathcal{Y}_{\Delta B}$, is
\bea 
\mathcal{Y}_{\Delta B} &\equiv& \frac{n_B}{s} = \frac{T^2}{6\,s} N_g \left( N_{u_L} + N_{u_R} + N_{d_L} + N_{d_R} \right) \,, \nonumber \\[1ex]
\mathcal{Y}_{\Delta B} &=& \frac{T^2}{2\,s} \left( \mu_{u_L} + \mu_{u_R} + \mu_{d_L} + \mu_{d_R} \right) \,,  \nonumber \\[1ex]
\mathcal{Y}_{\Delta B} &=& \frac{T^3}{2\,s} \left( 4\frac{\mu_{u_L}}{T} + 2 \frac{\mu_W}{T} \right) \ .   
\eea
The lepton number comoving asymmetry, $\mathcal{Y}_{\Delta L}$, is 
\bea
\mathcal{Y}_{\Delta L} & \equiv & \frac{n_L}{s} = \frac{T^2}{6\,s} N_g  \left(N_{e_L} + N_{e_R} + N_{\nu_L} \right)\,, \nonumber \\[1ex]
\mathcal{Y}_{\Delta L} & = & \frac{T^2}{2\,s} \left( \mu_{e_L} +\mu_{e_R} + \mu_{\nu_L} \right) \,, \nonumber \\[1ex]
\mathcal{Y}_{\Delta L} & = & \frac{T^3}{2\,s} \left( 3\frac{\mu_{\nu_L}}{T} + 2\frac{\mu_W}{T} - \frac{\mu_0}{T} \right) \, . 
\eea 

\subsubsection{Above the critical temperature for electroweak symmetry breaking}
\label{sec:aboveTC}
Above the critical temperature, $T_C$, for the breakdown of electroweak symmetry, $\mathcal{T}^3$ must vanish, so from Eq.~\eqref{eq:T3}, one concludes $\mu_W=0$. At $T>T_C$, a final relationship between chemical potentials is provided by ``sphaleron'' interactions, which take fields $d_Ld_Lu_L\nu_L$ from each generation into the vacuum; hence for each generation 
\be
2\mu_{d_L} + \mu_{u_L} + \mu_{\nu_L} = 2\mu_W + 3\mu_{u_L} + \mu_{\nu_L} = 3\mu_{u_L} + \mu_{\nu_L} = 0  , 
\label{eq:sphaleron}
\ee
where in the last expression we used $\mu_W=0$.  Combining this result with the result $Q=0$ (again with $\mu_W=0$) yields
\be
\mu_{u_L} = -\frac{7}{12} \mu_0 \ .  
\ee
Therefore, above $T_C$
\bea
\mathcal{Y}_{\Delta B} & = & 2\,\frac{T^3}{s} \, \frac{\mu_{u_L}}{T}
= - \frac{7}{6}\, \frac{T^3}{s} \, \frac{\mu_0}{T} \ ,
\nonumber \\[1ex]
\mathcal{Y}_{\Delta L} & = & \frac{T^3}{2s} \, \left( 3\frac{\mu_{\nu_L}}{T} - \frac{\mu_0}{T} \right) = - \frac{51}{14}\,\frac{T^3}{s} \, \frac{\mu_{u_L}}{T} = \frac{51}{24}\,\frac{T^3}{s} \, \frac{\mu_0}{T} \ .
\eea

Thus, an asymmetry in $\phi$ will convert to a baryon and a lepton asymmetry.

\subsubsection{Below the critical temperature for electroweak symmetry breaking}
\label{sec:belowTC}

After EWSB, the chemical potential for the fields inside an electroweak multiplet are no longer equal, so $\mu_{u_L} \neq \mu_{d_L}$ and $\mu_{e_L} \neq \mu_{\nu_L}$, the particles may no longer be relativistic, and there is a temperature dependence to the sphaleron process. This affects our calculations in Sec.\ \ref{sec:aboveTC}. However, this will only change the numerical factors of $\mathcal{Y}_{\Delta B}$ and $\mathcal{Y}_{\Delta L}$ and does not cancel the asymmetry produced.

\section{Conclusion and outlook}
\label{sec:conclusions}

Scalar fields which have non-minimal couplings to gravity are well-motivated inflaton candidates. Paradigmatic examples are the Higgs-inflation \cite{Bezrukov:2007ep} and $s$-inflation models \cite{Enqvist:2014zqa}.  In this paper, we have considered a scenario where several non-minimally coupled scalars contribute to the inflationary dynamics. In particular we investigated a model where these scalars are electroweak doublets and therefore generalize  Higgs inflation.  We focused on a setting where the dominant non-minimal coupling is allowed to be complex and investigated the effect that this would have on CP violation in our Universe.  We determined the inflationary dynamics in the regime where the model essentially conforms to the predictions of a single-field inflation model. The essential difference is that the inflaton obtains a non-zero phase representing a possible source of CP violation for subsequent post-inflationary evolution.  At the end of inflation, the inflaton particle, which is naturally assumed to have couplings with the SM Higgs, dumps its energy into the SM particle bath through the process of reheating, which populates the Universe with the SM particles. We showed how the complex value of the inflaton field leads to an asymmetry in the active scalar fields, and how this asymmetry will further be transmitted to the fermion sector.

This paper is intended to be a proof of concept that baryogenesis is viable in a CP-violating inflation scenario which requires a 3HDM setup.  As there are many unknown coupling constants and masses, a detailed model is not warranted at this time.  

We now review the assumptions and simplifications made in the analysis.
\begin{enumerate}
\item A three-Higgs doublet model: Three Higgs doublets are essential because the one- or two-Higgs doublet model cannot incorporate the CP-violating inflation phenomenon.
\item An inert and an active sector: Two Higgs doublets form inert doublets and are assumed to be heavier than the active doublet that serves as the SM Higgs doublet. The scalars in the inert sector will serve as inflaton candidates, while the Higgs in the active sector will serve as the SM Higgs.
\item Introduction of a $Z_2$ symmetry: This forbids FCNCs, and CP violation is only through the two heavy inflaton doublets.  Such CP violation does not contribute to EDMs.
\item Nonminimal couplings $\xi_1=\xi_2=0$: This assumption simplifies the analysis and allows us to focus on $\xi_4$ and its associated phase.
\item A proportional solution relating the imaginary part of the first doublet, $\eta_1$, and the real part of the second doublet, $h_2$, to the real part of the first doublet, $h_1$: This is for convenience; otherwise, one would have to analyze a three-field inflaton model. 
\item Immediate reheating: For immediate reheating we must have, at the end of inflation, the rate of $\phi$ production, $\Gamma_{\phi_1}=n_{\phi_1}\langle\sigma_{\phi_1\phi_1\to\phi\phi}|v|\rangle$, approximately equal to the expansion rate at the end of inflation, $H_e\approx \mu_1 h_1(\At)/\MPl$. We may approximate $\langle\sigma_{\phi_1\phi_1\to\phi\phi}|v|\rangle \approx |\lambda_3|^2/\mu_1^2$. A final approximation is to take at the end of inflation $n_{\phi_1}=\rho_{\phi_1}/\mu_1 = H_e^2\MPl^2/\mu_1$.  Together these approximations yield
\be
\frac{\Gamma_{\phi_1}}{H_e} \approx \frac{|\lambda_3|^2}{\sqrt{|\xi_4|}}\, \frac{\MPl^2}{\mu_1^2} \,. 
\ee
The requirement $\Gamma_{\phi_1}/H_e\approx 1$ can be easily satisfied.   If the ratio is much larger than unity we would be in the warm inflation region; if much less than unity, $\rho_{\phi_1}$ would decrease as the scale factor cubed.
\end{enumerate}

\subsubsection*{Acknowledgements}
VK acknowledges financial support from the Research Ireland Awards Grant 21/PATH-S/9475 (MOREHIGGS) under the SFI-IRC Pathway Program.  The work of EWK was supported in part by the US Department of Energy contract DE-FG02-13ER41958 and the Kavli Institute for Cosmological Physics at the University of Chicago.  EWK would like to thank the Dublin Institute for Advanced Studies for hosting a visit where this work was initiated. 

\bibliographystyle{JHEP} 
\bibliography{RockStar-Baryogenesis-references.bib}

\end{document}